\documentstyle[prb,preprint,aps,psfig]{revtex}
\begin{document}

\title{Spontaneous Coherence and Collective Modes
in Double-Layer Quantum Dot Systems}
\author{Jun Hu and E. Dagotto}
\address{National High Magnetic Field Laboratory, Florida State University,
1800 E. Paul Dirac Dr., Tallahassee, FL 32306, USA}
\author{A. H. MacDonald}
\address{Department of Physics, Indiana University,
Bloomington, IN 47405, USA}

\maketitle

\begin{abstract}

We study the ground state and the collective excitations of 
parabolically-confined double-layer quantum dot systems in a strong
magnetic field.  We identify parameter regimes where electrons form 
maximum density droplet states, quantum-dot 
analogs of the incompressible states of the bulk 
integer quantum Hall effect.  
In these regimes the Hartree-Fock approximation and the 
time-dependent Hartree-Fock approximations can be used to 
describe the ground state and collective excitations 
respectively.   We comment on the relationship between 
edge excitations of dots and edge magneto-plasmon 
excitations of bulk double-layer systems.

\end{abstract}
 
\pacs{68.65.+g} 

\section{Introduction}

In recent years, a large body of work has been denoted to the study  
of double-layer two-dimensional
electron systems~\cite{doublelayer} and also to
the understanding of electronic properties of two-dimensional quantum dot systems.~\cite{qdot}
In both cases such a huge interest has been triggered by 
advances in nanofabrication technology which made possible the
synthesis of these artificial systems.
In the case of double-layer systems,\cite{girvin1} some of the most interesting
novel physics has been uncovered in the strong magnetic 
field limit of quantum Hall effect.  Recent experimental~\cite{murphy} and 
theoretical\cite{fertig1,wen1,macdonald3,yang1,moon1} 
work in such strong magnetic fields 
has identified an unusual type of broken symmetry 
responsible for an unexpected quantum Hall effect which occurs 
at the Landau level filling factor of the individual layers
$\nu = \nu_T/2 = 1/2$.
This broken symmetry state has {\it spontaneous} interlayer
phase coherence, {\it i.e.} phase coherence even
in the absence of tunneling between the quantum wells.  
This effect is produced by the Coulomb interaction between electrons
in different layers.
For quantum dot systems it is also true that most of the 
research in this context has been focused on electronic properties in a strong magnetic
field.\cite{aspec}  Much of the work on quantum dots in this regime 
is related to the existence of {\it maximum-density-droplet} 
(MDD) states,\cite{ajp,pd} which are the quantum dot analogs
of the incompressible states responsible for the quantum Hall
effect in bulk systems, and to the edge reconstructions\cite{erecons}
which occur when these states become unstable.  The recent experimental
realization\cite{lqdots} of layered quantum dot systems adds to
the motivation for theoretical studies of these systems.
In this article we discuss MDD states 
of double-layer quantum dot systems.  Complementary numerical
exact diagonalization study of double-layer quantum dot systems
in a strong magnetic field have appeared recently~\cite{palacios} 
and some preliminary results\cite{confpaper} from the present study have been
reported earlier.  

The paper is organized as follows.  In Section II the model used here
for two-dimensional double-layer quantum
dot systems is presented and discussed. In Section III the stability limits for the 
$\nu_T =1$ MDD state are analyzed.  This state is the quantum
dot analog of the phase-coherent incompressible state in the bulk
limit.
In Section IV we discuss collective excitations of this state
with an emphasis on the interplay between the gapless edge excitations,
always\cite{edge} associated with the quantum Hall effect, and 
the gapless Goldstone modes of the broken-symmetry ground state
in the bulk.  In Section V We turn our attention to the $\nu_T=2$ MDD state
,which corresponds to the bulk state with a 
filled Landau level in each layer.  Our emphasis here is
on the relationship between low energy excitations of the 
quantum dot system and coupled edge magneto-plasmon modes of 
bulk double-layer systems.  Our results are briefly summarized in
Section VI.  

\section{The Model}

We model a double-layer system made out of  two-dimensional 
quantum dots by assuming identical parabolic potentials, 
$V(r) = \frac{1}{2}m^*\Omega^2 r^2$, in the two layers.
(In Section V it is  emphasized 
that many of our results apply equally well to systems with 
bulk two-dimensional electrons in each layer.) 
The system is placed in a strong
magnetic field $\vec B$ perpendicular to the layers.  We restrict our attention
to the strong magnetic field limit where 
$\Omega/\omega_c << 1$, and only the states in the lowest Landau level are 
important.  (Here $\omega_c = eB/m^*c$ is the cyclotron frequency.)  The
lowest-Landau-level single-particle eigenstates in the symmetric
gauge\cite{qdot} are labeled by the angular momentum $m$:
\begin{equation}
\langle \vec r|m \rangle = \frac{1}{\sqrt{2\pi
\ell^2 2^m m!}}\left(\frac{z}{\ell}\right)^m\exp
\left(-\frac{|z|^2}{4\ell^2}\right)
\end{equation}
and 
\begin{equation}
\varepsilon_m = \frac{1}{2}\hbar \omega_c + \gamma (m+1)
\end{equation} 
where $\gamma = m^*\Omega^2\ell^2 = \hbar \omega_c (\Omega/\omega_c)^2 $,
$\ell^2 \equiv \hbar c / eB $, $ \vec r = (x,y)$, $z= x+iy$ is the 2D 
electron coordinate expressed as a complex number, and the allowed
values of the single-particle angular momentum within the lowest Landau level 
are $m = 0, 1, 2, ...$.  We will assume that the
electron system is completely spin-polarized by the magnetic field.

Including a phenomenological term describing tunneling
between the two quantum dots and
up to an irrelevant constant,  
the second-quantized 
Hamiltonian of the system is given by 
\begin{eqnarray}
 \cal H & = & \sum_{m \sigma}m\gamma c_{m\sigma}^{\dag} c_{m\sigma} -
\sum_{m\sigma\sigma^\prime}(1-\delta_{\sigma\sigma^\prime})
tc_{m\sigma}^{\dag}c_{m\sigma^\prime} + \nonumber \\ 
& & + \frac{1}{2}\sum_{\begin{array}{c}\vspace{-0.33in} {\scriptstyle m_1m_2 m_1^\prime m_2^\prime }\\[-0.15in] 
{\scriptstyle \sigma\sigma^\prime}\end{array}}
V_{m_1^\prime m_2^\prime m_1 m_2}^{\sigma\sigma^\prime}
c^{\dag}_{m_1^\prime\sigma}c^{\dag}_{m_2^\prime\sigma^\prime}
c_{m_2\sigma^\prime}c_{m_1\sigma} 
\end{eqnarray}
where 
\begin{equation}
V_{m_1^\prime m_2^\prime m_1 m_2}^{\sigma\sigma^\prime} =
\langle m_1^\prime m_2^\prime|V_0|m_1m_2 \rangle +
\sigma\sigma^\prime \langle m_1^\prime m_2^\prime|V_z|m_1m_2 \rangle ,
\label{eq:v0vz}
\end{equation}
and $t$ is the hopping amplitude between the 
two layers. The layer index is $\sigma = \uparrow, 
\downarrow$ where invoking a helpful\cite{macdonald2} analogy between the
layer degree of freedom and the spin degree of freedom, 
$\sigma = \uparrow$ corresponds to electrons in the right layer and 
$\sigma = \downarrow$ to electrons in the left layer.
In Eq.~\ref{eq:v0vz}, $V_0 = (V_A+V_E)/2 $ and $V_z = (V_A-V_E)/2$,
are proportional to the sum and difference of intra-layer ($V_A$) 
and interlayer ($V_E$) Coulomb interactions.

For $\nu_T =1$ the 
ground state of the quantum dot 
in the Hartree-Fock approximation 
has its pseudospin polarized in the $\hat x-\hat y$ 
pseudospin plane.  Physically, electrons in this state occupy
states which are a coherent linear combination of states localized
in the separate layers.  In order to study this state it is 
convenient to transform to a new representation,
carrying out a rotation in pseudospin space by defining
\begin{equation}
c_{m\sigma}^\dagger =
\frac{1}{\sqrt{2}}(\sigma\alpha_{m\sigma}^\dagger+\alpha_{m\bar
\sigma}^\dagger).
\end{equation}
In this representation, a pseudospin up electron is in a 
symmetric double-layer state and a pseudospin down electron is in 
an antisymmetric state.  The Hamiltonian can alternately be expressed 
in the form: 
\begin{eqnarray}
\cal H & = & \sum_{m \sigma}(m\gamma -\sigma t)\alpha^\dagger_{m\sigma}
\alpha_{ m\sigma}+ \nonumber \\
 & &  \frac{1}{2}\sum_{m_1m_2m_3m_4\sigma\sigma^\prime} (\langle m_1 m_2|V_0|m_3 m_4 \rangle
\alpha^\dagger_{m_1\sigma}\alpha^\dagger_{m_2 \sigma^\prime}
\alpha_{m_3\sigma^\prime}\alpha_{m_4\sigma} \nonumber \\
 & &  \hspace*{0.9in} + \langle m_1 m_2|V_z| m_3 m_4 \rangle
\alpha^\dagger_{m_1\sigma}\alpha^\dagger_{m_2 \sigma^\prime}
\alpha_{m_3\bar \sigma^\prime}\alpha_{m_4\bar \sigma}),
\label{eq:hamsas}
\end{eqnarray}
where $\bar \sigma = \downarrow,\uparrow$ for $\sigma =\uparrow,
\downarrow$.
In this rotated pseudospin representation the hopping parameter
$t$ simply acts as an external magnetic field. 
We will discuss here only the limit $t \rightarrow 0$.

\section{The Stability of MDD state}

For $N$ non-interacting electrons in a single-layer
quantum dot, the many-body ground state
is a single Slater determinant in which the confinement energy
is minimized by occupying the orbitals from $m=0 $ to $m=N-1$. This state
is an exact many-body eigenstate of the Hamiltonian even when
electron-electron interactions are included.~\cite{ajp} 
We refer to
this state as the maximum density droplet (MDD) state.
The MDD state is the analog for quantum dots of the bulk
$\nu =1$ state.  For double layers a new type of MDD state 
can occur which is the analog of the bulk double-layer $\nu_T=1$ 
broken-symmetry\cite{bsapology} ground state.\cite{fertig1,wen1,yang1,moon1}
In this state electrons occupy symmetric states
(pseudospin up) with $m=0$ to $m=N-1$ such that the total angular 
momentum is $M_0 = N (N-1)/2$.
Many-body states with smaller total angular momentum than this MDD state
will have the advantage of smaller confinement energies
since the electrons are closer to the minimum of the confinement
potential, but the disadvantage of larger interaction energies since 
the electrons
are closer to each other.  For weak confinement the
angular momentum of the ground state of an interacting-electron droplet
will occur at $M > M_0$ 
in order to reduce the Coulomb energy while for 
sufficiently strong confinement the ground state will occur at $M < M_0$
in order to reduce the confinement energy. 

A necessary condition for the MDD state to be the ground state is that 
all occupied orbitals have a Hartree-Fock quasiparticle energy 
which is lower than the quasiparticle energies of all unoccupied
orbitals.  Using Eq.~(\ref{eq:hamsas}) we have found that 
the Hartree-Fock quasiparticle energies for the symmetric 
MDD state are given by 
\begin{equation}
\varepsilon_{m\sigma}^{HF} = m\gamma-\sigma t + \sum_{m^\prime}
n_{m^\prime\sigma}U_{mm^\prime}^0+\sum_{m^\prime}n_{m^\prime\bar
\sigma}U_{mm^\prime}^z
\end{equation}
where
\begin{equation}
        U^0_{mm^\prime} = \langle mm^\prime|V_0|mm^\prime \rangle - 
\langle m^\prime m|V_0|mm^\prime \rangle
,
\end{equation}
\begin{equation}
        U^z_{mm^\prime} = \langle mm^\prime|V_0|mm^\prime \rangle - 
    \langle m^\prime m|V_z|mm^\prime \rangle.
\end{equation}
For the symmetric MDD state $n_{m\sigma}$ is $1$ for $\sigma = \uparrow$
and $m < N$, and it is zero otherwise.

\begin{figure}
\centerline{\psfig{figure=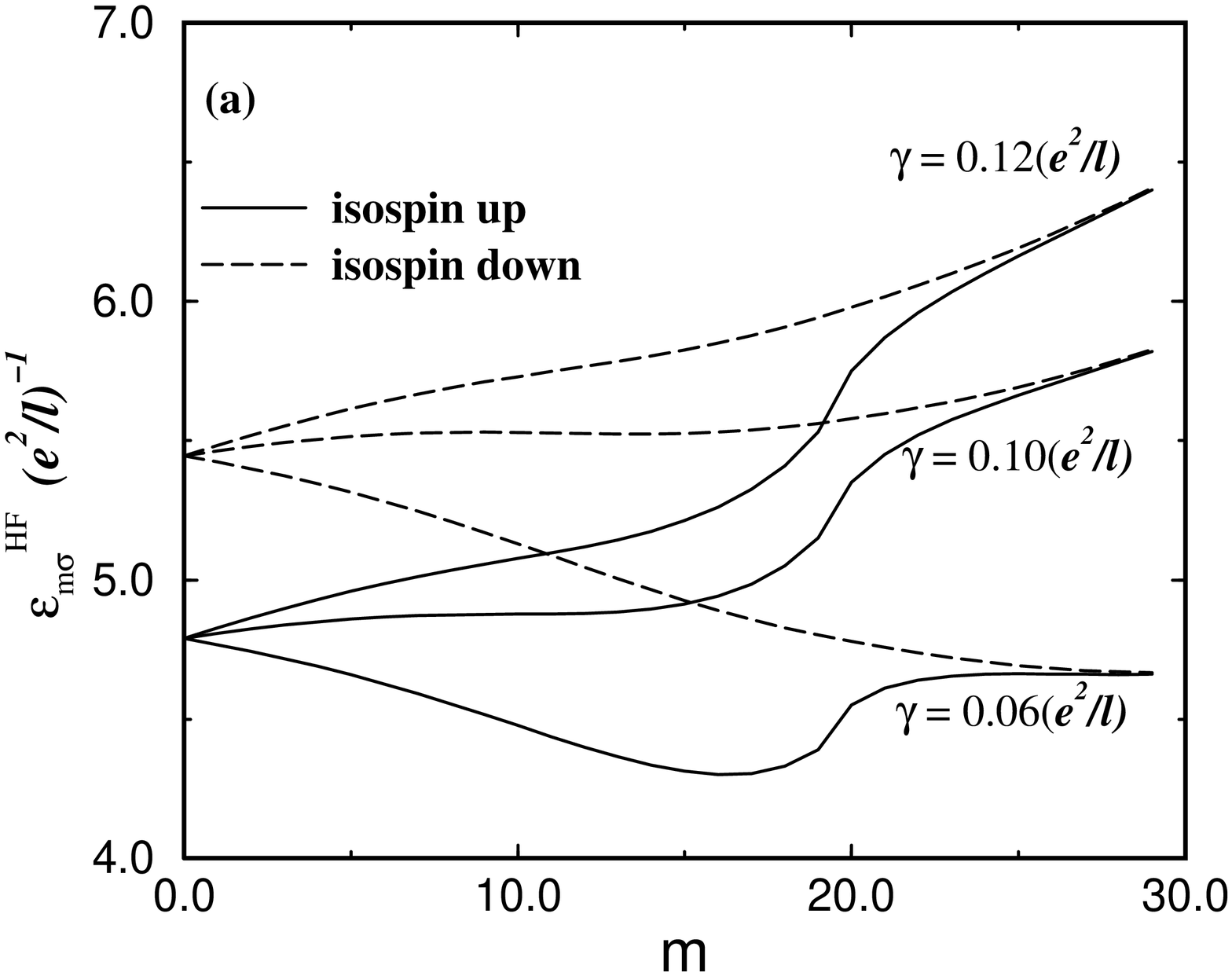,width=8cm,angle=270}
\psfig{figure=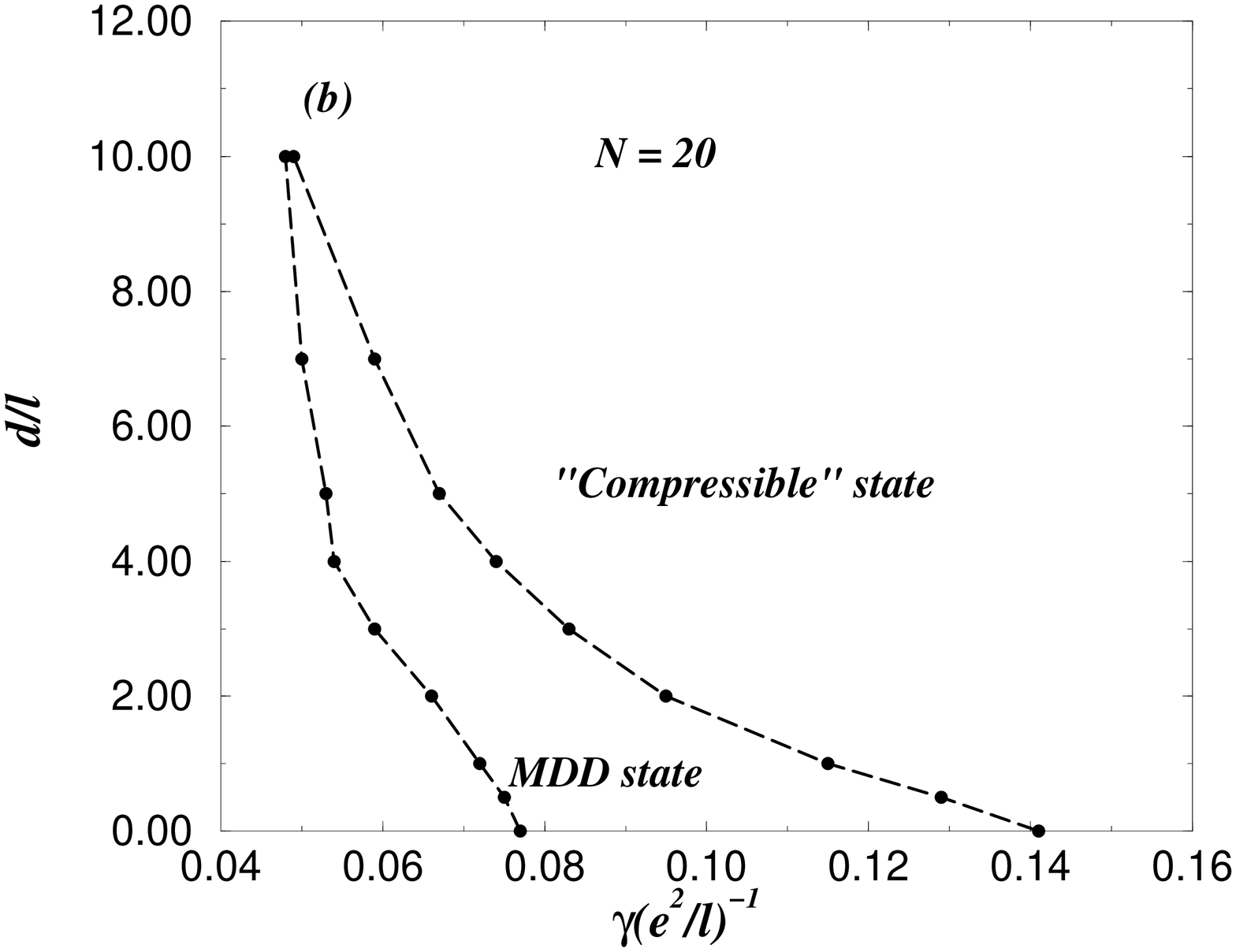,width=8cm,angle=270}}
\caption{(a) The single particle H-F energy for the double-layer quantum
system with 20 electrons and interlayer distance $d = \ell$ at different
confinement potential strengths.
(b) The phase diagram for MDD states in $N=20$ double-layer quantum
dots. The ground state in the region of "compressible" state tends to
have many low energy excitation away from the edge of quantum dots. }
\label{fig1}
\end{figure}

Fig.~\ref{fig1}a shows Hartree-Fock quasiparticle energies for 
symmetric and antisymmetric orbitals for $N=20$, $t=0$, and  $d = \ell$ at 
three different confinement strengths.
The exchange splitting between the occupied symmetric orbitals
and the unoccupied antisymmetric orbitals is due to interlayer 
interactions and decreases in magnitude as $d$ increases or as 
$m$ increases toward the edge of the dot.  
For $\gamma/(e^2/\epsilon \ell) = 0.06$ the confinement potential is relatively
weak and unoccupied symmetric orbitals near the edge of the dot are lower in
energy than occupied symmetric orbitals near the center of the dot.
For $\gamma/(e^2/\epsilon \ell) =0.12$ the confinement potential
is now strong and unoccupied antisymmetric orbitals near the center
of the dot are lower in energy than occupied symmetric orbitals
near the edge of the dot.  
These results show that there is a finite  interval of confinement 
strengths, including the case $\gamma / (e^2/\epsilon \ell) = 0.10$,
for which the MDD state is stable.  Fig.~\ref{fig1}b 
shows a phase diagram constructed for $N=20$ using this quasiparticle energy
stability criterion, which shows how the interval  where the
confinement potential strength renders the MDD state 
stable narrows and shifts as a function of layer separation.
(The actual stability region for the 
MDD state beyond the Hartree-Fock approximation will be slightly
narrower because of excitonic corrections to particle-hole
excitation energies).  For $d \to 0$ interlayer and intralayer 
interactions become identical and these results would apply equally 
well to the $\nu_T=1$ state for a single-layer system in 
the limit of vanishing Zeeman coupling (Note that we assume {\it strong}
Zeeman coupling for the double-layer systems discussed in this paper). 
For larger $N$ the region of stability of a parabolically confined 
MDD droplet becomes smaller; for very large $N$ the electron density 
in a parabolic external potential must follow the semi-elliptic
behavior prefered\cite{glazmanes} by electrostatics rather than
the nearly constant charge density profile of the MDD state.  For  
sufficiently large $N$ the MDD state is never stable in a parabolic external
potential and can be stabilized only by the potential from an
approximately neutralizing positively charged background.  

\section{Collective Excitations}

In the time-dependent Hartree-Fock approximation, elementary 
excitations of the MDD state are constructed by allowing its
particle-hole excitations to couple.  The Hamiltonian for the 
double-layer system does not mix states in which the 
number of particles in symmetric (pseudospin $\uparrow$) states
differs by an odd integer; elementary excitations with even
and odd numbers of pseudospin reversed states do not mix.  In
order to describe the odd excitations we introduce the 
operators 

\begin{equation}
\rho_{M,m}^- = \alpha^+_{m+M,\downarrow}\alpha_{m,\uparrow} \mbox{   and   }
\rho_{-M,m}^+ = \alpha^+_{m,\uparrow}\alpha_{m-M,\downarrow}.
\end{equation}
The equations of motion for these operators 
are readily derived and 
reduce to a closed system of equations
after Hartree-Fock factorization: 
\begin{equation}
\left( \begin{array}{c}
    {[{\cal H}, \rho_{M,m}^{-}]} \\ {[{\cal H}, \rho_{-M,m}^{+} ]}
        \end{array} \right)
     = \left( \begin{array}{cc}
         E_{mm^\prime}^M & F_{mm^\prime}^M \\
        -F_{mm^\prime}^{-M} & -E_{mm^\prime}^{-M}
                                \end{array} \right)
                        \left( \begin{array}{c}
                        \rho_{M,m^\prime}^- \\
                        \rho_{-M,m^\prime}^+
                                \end{array}
                        \right)
\label{eq:odd}
\end{equation}
where
\begin{eqnarray}
        E_{mm^\prime}^M & = &
\delta_{mm^\prime}(\varepsilon_{m+M\downarrow}^{HF
} - \varepsilon_{m\uparrow}^{HF}) + \nonumber \\
& & (\langle m,m^\prime+M|V_z|m+M,m^\prime \rangle - \langle m^\prime+M,m|V_0|m+M,m^\prime \rangle )
\label{eq13}
\end{eqnarray}
and
\begin{equation}
        F_{mm^\prime}^M = \langle m m^\prime|V_z|m+M,m^\prime-M \rangle - 
\langle m^\prime m |V_z|m+M, m^\prime - M \rangle .
\label{eq14}
\end{equation}
Diagonalizing the above matrix will give us the time-dependent 
Hartree-Fock approximation (TDHFA) 
pseudospin-flip elementary excitations of 
the MDD state.  If these operators evolved according to the 
mean-field Hartree-Fock Hamiltonian for the ground state,
the excitation energies would simply equal the difference
of occupied and unoccupied Hartree-Fock eigenvalues.  The 
additional terms in Eq.~(\ref{eq:odd}) reflect the changes 
in the mean-field Hamiltonian in the excited state and, in 
a diagrammatic derivation of the TDHFA
, would arise from vertex and repeated-bubble
corrections to the single-loop approximation for the pseudospin-flip
response function.  Below we refer to the influence of these
additional terms as vertex corrections.

The even elementary excitations are found by considering the 
equation of motion for the operator
$\rho_{M,m}^0 = \alpha_{m+M,\uparrow}^\dagger\alpha_{m,\uparrow} $.
Hartree-Fock factorization of the equation of
motion of these operators again gives a closed set of equations:
\begin{equation}
[{\cal H}, \rho_{M,m}^{0}] = G_{mm^\prime}^M \rho_{M,m^\prime}^0.
\label{eq:even}
\end{equation}
Diagonalizing the matrix $G_{mm^\prime}^M $ gives the even 
elementary excitations of the MDD state.  In Eq.~(\ref{eq:even}) 
\begin{eqnarray}
        G_{mm^\prime}^M  & = & \delta_{mm^\prime} 
       ( \varepsilon_{m+M\uparrow}^{HF} - \varepsilon_{m\uparrow}^{HF}
	 )
\nonumber
 \\
&+ & (\langle mm^\prime+M|V_0|m+Mm^\prime \rangle - \langle m^\prime+M m|V_0|m+M m^\prime \rangle).
\label{eq16}
\end{eqnarray}
Again vertex corrections cause 
the excitation energies to differ from the difference of 
Hartree-Fock eigenvalues.

\begin{figure}
\centerline{\psfig{figure=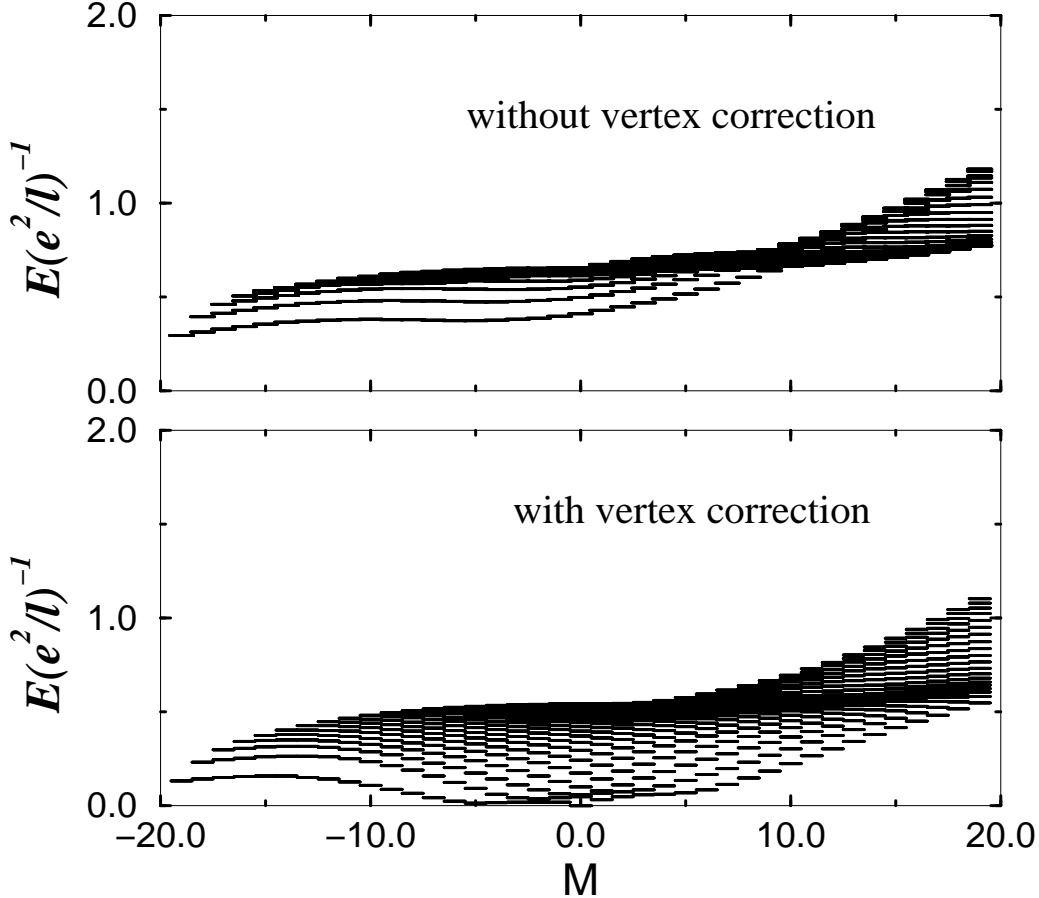,width=15cm,angle=0}}
\caption{Isospin-flip excitations at $\nu_{T} =1$ 
with vertex correction and without vertex correction
for a double-layer quantum dot system with $N = 20$ electrons,  
interlayer distance $d = \ell$, and
confinement potential $\gamma = 0.10 e^2/\ell$.}
\label{fig2}
\end{figure}

Fig.~\ref{fig2} shows the odd elementary excitations
obtained for layer separation $d = \ell$, $N=20$, 
and confinement potential $\gamma = 0.10 (e^2/ \ell)$ as a 
function of angular momentum $M$.
Note that for each $M$ the excitation energy has a contribution
$M \gamma$ from the confinement potential.
The number of particle-hole excitations with angular momentum change
$M$ is $N$ for $M \ge 0$ and $N + M$ for $M < 0$.  (The minority-spin
angular momentum cannot be negative.)  When vertex corrections
are neglected, the lowest excitation energies occur 
at the most negative angular momentum and correspond to 
transferring an electron from pseudospin up at the edge of the 
quantum dot to pseudospin down at the center of the quantum dot.
Vertex corrections change the elementary
excitation spectrum qualitatively.  The most obvious modification is the 
appearance of a zero-energy excitation for $M=0$ which 
we discuss below.  The $ M = - N + 1$ 
excitation energy is also reduced by vertex corrections;  
with increasing $\gamma$ the energy of this excitation will become
negative before the quasiparticle energy stability criterion 
is violated so that the region of stability of the MDD
state is narrower than implied by this criterion.  
A similar reduction in the excitation energies occurs at 
positive $M$.   

In bulk systems the broken symmetry ground state of the 
$\nu_T =1$ double-layer system has a gapless Goldstone 
mode\cite{girvin1} with linear dispersion (It can be shown that
the $M=0$ zero-energy
elementary excitation just corresponds to a 
global rotation of the pseudospin). Some 
remnant of these Goldstone modes should be present in the 
excitation spectrum of finite-size quantum dot systems.  
In the limit of large dots, we expect that $M/R$ 
should act like the azimuthal component of a two-dimensional 
wave vector and that the many collective modes that occur at a 
given $M$ are approximately related to the discrete set 
of radial wavevectors geometrically defined by the finite 
radius $R_N \approx \sqrt{2N} \ell$ of the quantum dot which is
related, at least approximately, to the zeroes of $J_0(qR_N)$.  
We have, however, not yet been able to find a completely satisfactory  
semiclassical interpretation of our microscopic TDHFA 
results in this way.  For example the pronounced 
asymmetry between the low-energy 
TDHFA collective modes for $M < 0$ and $M > 0$. 

\begin{figure}
\centerline{\psfig{figure=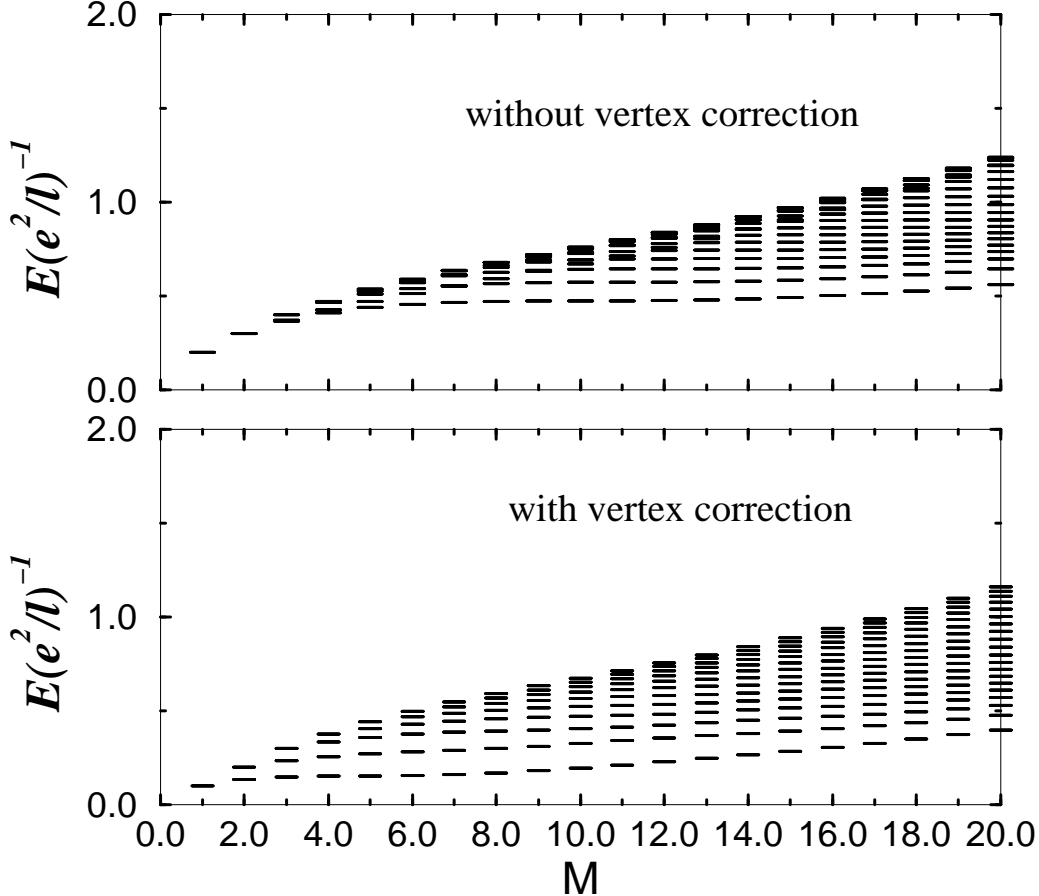,width=15cm,angle=0}}
\caption{Edge Charge density-wave edge excitations 
 with vertex correction and without vertex correction
at $\nu_{T} =1$ 
for a double-layer quantum dot system with $N = 20$ electrons and
interlayer distance $d = \ell$ and
confinement potential $\gamma = 0.10(e^2/\ell)$.  }
\label{fig3}
\end{figure}

The even elementary excitations represent charge-density-wave 
edge excitations of the incompressible MDD state and
occur only for $M > 0$.  In Fig.~\ref{fig3} we show TDHFA 
results calculated for $N=20$, $\gamma = 0.1 (e^2 / \epsilon \ell)$,
and $d = \ell$.  Again the excitation energies are substantially
reduced by vertex corrections.  These results are identical to those that
would be obtained for a single-layer system with an interaction
which is the average of the double-layer intra-layer and 
inter-layer interactions.  As discussed in detail previously for 
the single-layer case,\cite{ajp} the $M=1$ excitation energy is exactly
equal to $\gamma$ and is not influenced by electron-electron
interactions.  These charge density wave excitations of the 
quantum dot are harbingers of the edge magneto-plasmon excitations 
of bulk systems which we discuss at greater length in the 
following section.  With decreasing confinement strength a 
collective mode will become unstable slightly before the 
quasiparticle-energy stability criterion is violated, narrowing 
the region of stability of the MDD state on the weak confinement
side just as it is narrowed on the strong confinement side. 
Very recently\cite{sondhikarlhede} for the $d=0$ case, Sondhi {\it et al.} have 
identified topological edge instabilities 
which further limit the stability region of MDD states. 

\section{Edge magneto-plasmons and $\nu_T=2$ Double-layer Quantum Dots}

The quantum dot analog of the bulk $\nu=2$ state in a double-layer system
is the single Slater determinant state
in which the single-particle angular momentum states 
from $m=0$ to $m=N-1$ are occupied in each layer.  It can be 
shown that this state is a pseudospin singlet.  In quantum dot
systems this state becomes unstable with increasing confinement
strength only when it becomes energetically favorable to occupy
higher Landau levels, and becomes unstable with 
decreasing confinement strengths when it becomes favorable to move
electrons away from the center of the quantum dot, for example
by edge reconstruction.  For parabolic confinement, the 
region of stability depends on $d$ and narrows with increasing 
$N$.  The excitation spectrum of this state is less subtle
than for the $\nu =1$ case discussed above because a gap 
$\sim \hbar \omega_c$ exists in the bulk for both pseudospin-flip
{\it and} charge density excitations.  The only low-energy excitations 
are localized at the edge of the system.

In this section we work 
in the representation where pseudospin eigenstates are localized
in individual layers.  We consider the 
equations of motion for the operators which create 
in phase and out-of-phase density wave excitations at 
the edges of the two-layers:
\begin{equation}
\rho^{o}_{m,M} = \sum_{\sigma} c^{\dagger}_{m+M,\sigma} c_{m,\sigma}
\label{eq:inphase}
\end{equation}
and 
\begin{equation}
\rho^{z}_{m,M} = \sum_{\sigma} \sigma c^{\dagger}_{m+M,\sigma} c_{m,\sigma},
\label{eq:outofphase}
\end{equation}
where $m <  N $ and $m+M \ge N$.  
After a Hartree-Fock factorization the equations of motion for
these operators close and a calculation very similar to that 
detailed in Section IV gives the following result:
\begin{eqnarray}
[{\cal H}, \rho^{0}_{m,M}]  &=&
(\epsilon^{HF}_{m+M} - \epsilon^{HF}_m) \rho^{0}_{m,M} \\
& + &  \sum_{m'} \rho^{0}_{m',M}
\big[ 2  \langle m, m'+M | V_0 | m+M, m' \rangle 
- \langle m,m'+ M | V_A | m', m+M \rangle \big]
\label{eq:inphaseeng}
\end{eqnarray}
and 
\begin{eqnarray}
[{\cal H}, \rho^{z}_{m,M} ] & = &
(\epsilon^{HF}_{m+M} - \epsilon^{HF}_m) \rho^{z}_{m,M}  \\
& + & \sum_{m'} \rho^{z}_{m',M}  
\big[ 2 \langle m,m'+M | V_z | m+M, m' \rangle 
 - \langle m,m'+M | V_A | m', m+M \rangle \big].
\label{eq:outofphaseeng}
\end{eqnarray} 
In this case the single-particle Hartree-Fock energies,
\begin{equation}
\epsilon^{HF}_m = \gamma (m+1) + \sum_{m'} 
\big[2  \langle m,m'|V_0|m,m'\rangle - \langle m,m' |V_A | m',m \rangle \big],
\label{eq:nu2hf}
\end{equation} 
are pseudospin independent.  

\begin{figure}
\centerline{\psfig{figure=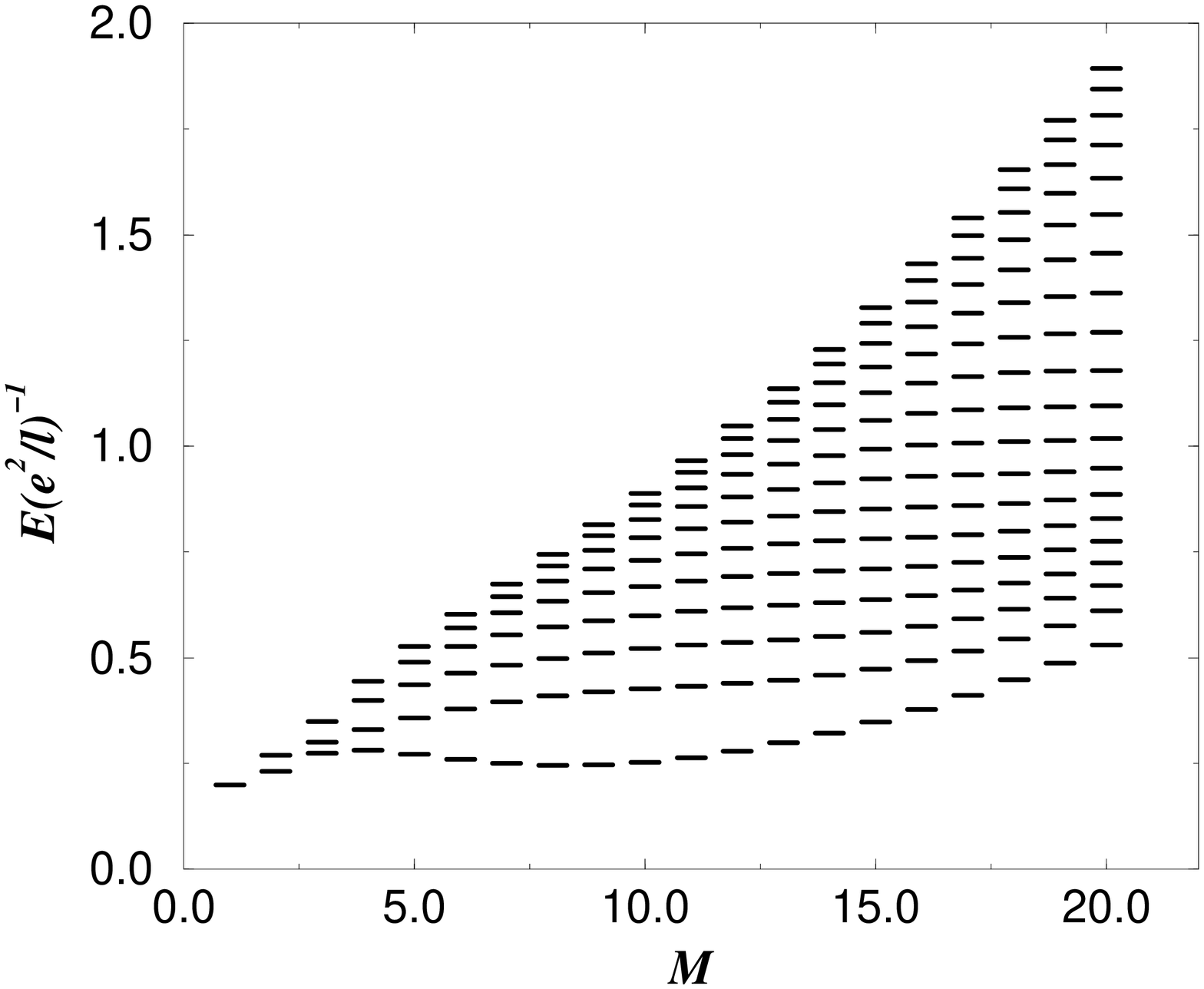,width=15cm,angle=270}}
\caption{Spin-density wave edge excitations at $\nu_{T} =2$ 
for a double-layer system  with $N = 20$,
interlayer distance $d = \ell$, and
confinement potential $\gamma = 0.20 (e^2/\ell)$.}
\label{fig4}
\end{figure}

\begin{figure}
\centerline{\psfig{figure=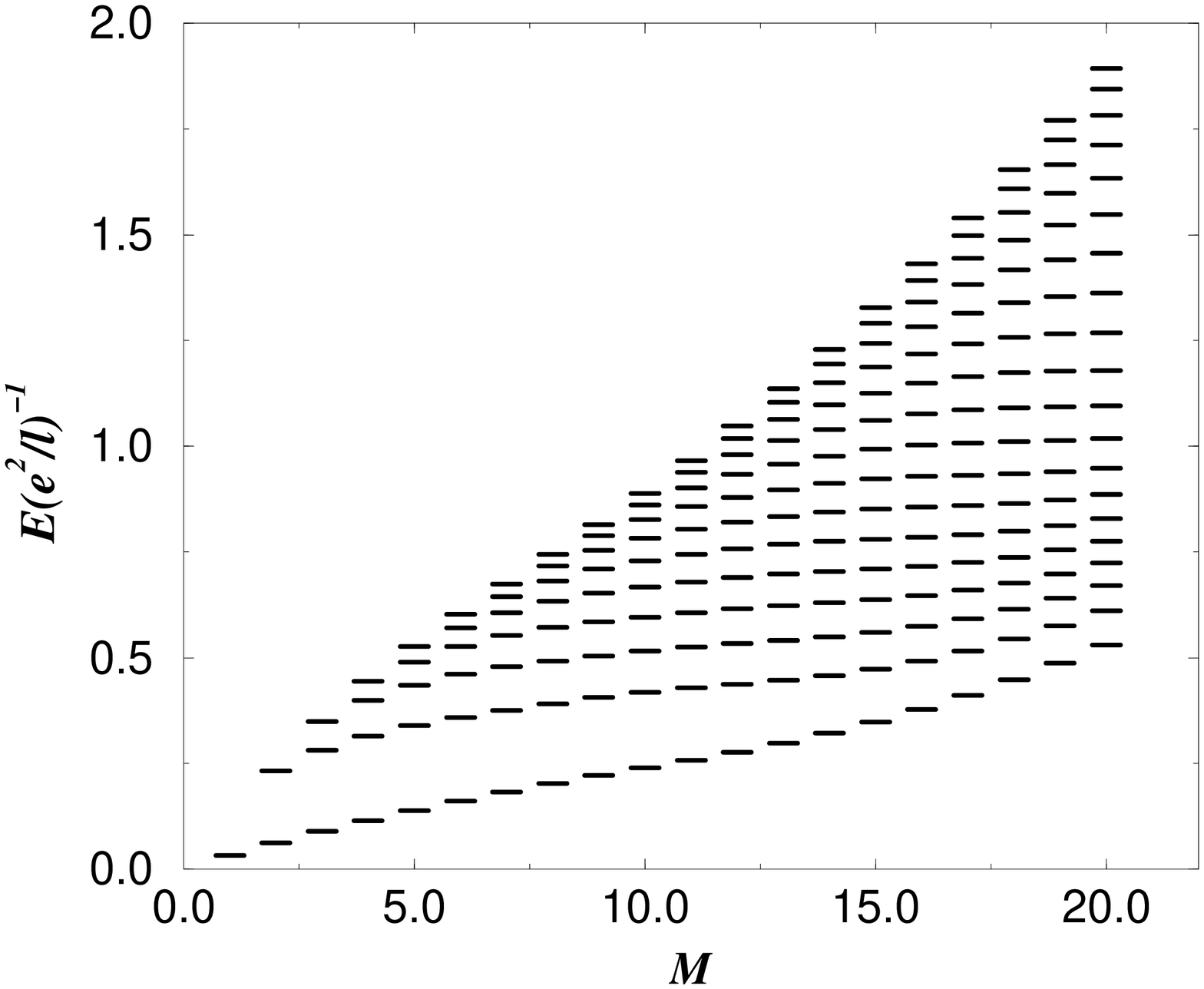,width=15cm,angle=270}}
\caption{Charge-density wave edge excitations at  
$\nu_{T} =2$ for a double-layer system with $N = 20$,
interlayer distance $d = \ell$, and
confinement potential $\gamma = 0.20 (e^2/\ell) $.}
\label{fig5}
\end{figure}

Illustrative TDHFA 
quantum dot collective excitation energies at $\nu_T=2$, 
evaluated using these expressions, are shown in 
Figs.~\ref{fig4} and ~\ref{fig5}.  We comment primarily on  
the $M=1$ results which we relate to edge magneto-plasmons 
of bulk double-layer systems below.  For $M=1$, 
$m = N-1$ in both Eq.~(\ref{eq:inphase}) and 
Eq.~(\ref{eq:outofphase}), explicit expressions can be 
given for the excitation energies:
\begin{eqnarray}
E^{(0)}_1 & = & \gamma + 
\sum_{m'=0}^{N-1} \big[2  \langle N,m'|V_0|N,m'\rangle -
 \langle N,m' |V_A | m',N \rangle \big]  \nonumber \\
 & & - \sum_{m'=0}^{N-1} \big[2  \langle N-1,m'|V_0|N-1,m'\rangle -
 \langle N-1,m' |V_A | m',N-1 \rangle \big] \nonumber \\
& & + 2 \langle N-1,N | V_0 |N, N-1 \rangle - 
 \langle N-1, N | V_A | N-1,N \rangle ,
\label{eq:engM1inphase}
\end{eqnarray} 
and 
\begin{eqnarray}
E^{(z)}_1 & = & \gamma + 
\sum_{m'=0}^{N-1} \big[2  \langle N,m'|V_0|N,m'\rangle -
 \langle N,m' |V_A | m',N \rangle \big]  \nonumber \\
 & & - \sum_{m'=0}^{N-1} \big[2  \langle N-1,m'|V_0|N-1,m'\rangle -
 \langle N-1,m' |V_A | m',N-1 \rangle \big] \nonumber \\
& & + 2 \langle N-1,N | V_z |N, N-1 \rangle - 
 \langle N-1, N | V_A | N-1,N \rangle. 
\label{eq:engM1outphase}
\end{eqnarray} 

The in-phase mode has interaction corrections from the 
difference of the interaction contribution to the Hartree-Fock 
single-particle energies at $m=N$ and $m=N-1$ and also from
vertex corrections.  In the case of single-layer quantum 
dots\cite{ajp} these two classes of corrections are known
to cancel exactly.  The easiest way to establish this property
in the single layer case 
is to note that the $M=1$ excitation operator simply changes the 
state of the center-of-mass of all electrons without 
changing the relative motion state to which interactions 
are sensitive.  For $E^{(0)}_1$ the same argument goes 
through in the double layer case and only the external
potential contribution to the excitation energy survives:
\begin{equation}
E^{(0)}_1 = \gamma.
\label{eq:eng1simp}
\end{equation}
This cancelation is verified numerically as seen in Fig.~\ref{fig4}.
Eq.~(\ref{eq:eng1simp}) can also be derived from the following 
identities which are established for interaction matrix elements
in the lowest Landau level:
\begin{equation}
\sum_{m'=0}^{N-1} \langle N,m'|V|N,m'\rangle 
 - \sum_{m'=0}^{N-1} \langle N-1,m'|V|N-1,m'\rangle 
 = -\langle N-1,N | V |N, N-1 \rangle 
\label{eq:jh1}
\end{equation}
and 
\begin{equation}
\sum_{m'=0}^{N-1} \langle N,m'|V|m',N\rangle -
  \sum_{m'=0}^{N-1} \langle N-1,m'|V|m',N-1\rangle 
  = -\langle N-1,N | V |N-1, N \rangle.
\label{eq:jh2}
\end{equation}
These identities hold for any interaction potential and 
in particular for $V=V_0$ or $V=V_A$. 
It follows from Eq.~(\ref{eq:jh1}) and Eq.~(\ref{eq:jh2}) that 
\begin{equation}
E^{(z)}_1 = \gamma - 2 \langle N-1,N | V_E | N, N-1 \rangle.
\label{eq:eng1outsimp}
\end{equation}
This simple result has also been verified numerically.

These $M=1$ edge excitations are the quantum dot precursors 
of the edge magnetoplasmon collective excitations\cite{empl} of larger $N$ 
electronic systems in strong magnetic fields which give rise
to far-infrared resonances in micron scale systems and 
to surprisingly sharp radio-frequency resonances in large area
2DEG's.  Some helpful remarks on edge magnetoplasmons 
in double-layer systems follow from the above analysis.
As we have emphasized, larger $N$ MDD states are bulk 
quantum Hall states and they will occur only when the external
potential is quite close to that from a neutralizing background 
of positive charge.  For a background
charge density which neutralizes the electron density in each layer
it follows from Eq.~(\ref{eq:jh1}) that the difference
between the expectation values of the external potential for the 
states with angular momenta $N$ and $N-1$ is 
\begin{equation}  
\tilde \gamma = \langle N-1,N | V_A + V_E  |N, N-1 \rangle
\approx  \frac{e^2}{\pi R} [ \ln (R/\ell) + \ln (R/d) ].
\label{eq:emplasmona}
\end{equation}
The final form for $\tilde \gamma$ assumes an electron disk 
whose radius $R$ is large compared to both the magnetic 
length $\ell$ and the layer separation $d$.   (We assume that 
$d$ is at least $\sim \ell$.)  The two terms in square brackets
in Eq.~(\ref{eq:emplasmona}) comes respectively from intra-layer 
and inter-layer 
interactions with the background.   As discussed above, the 
energy of the in-phase collective magnetoplasmon excitations of the
double-layer system ($E_{mpl}^{+}$) depends only on the external potential and 
is therefore given by $E_{mpl}^{+} =  \tilde \gamma$.   Since 
$ \langle N-1,N | V_E | N, N-1 \rangle \sim (e^2/ \pi R) \ln (R/d) $ 
for $R \gg d$ it follows that the energy of the out-of-phase mode is
\begin{equation}
E_{mpl}^{-} = \frac{e^2}{\pi R} [ \ln (R/\ell) - \ln (R/d) ].
\label{eq:emplasmonb}
\end{equation}
For large 2DEG's the splitting of the edge magnetoplasmon 
energies due to coupling between the layers persists to very 
large layer separations; for example $E_{mpl}^{+}=3 E_{mpl}^{-}$ 
for $ d = (R \ell)^{1/2}$.  When the separation between the two layers
is comparable to $\ell$, often a requirement for observable  
coupling effects in bulk systems, the edge magnetoplasmon
coupling will typically make $E_{mpl}^{-}$ unobservably small.  
We remark that these results have been derived for disorder free
systems.  We expect that these results for edge magnetoplasmon 
energies will remain valid in disordered systems provided that 
the microscopic length scale $\ell$, which describes the degree
of localization of the edge wave in the clean limit, is replaced
by the appropriate\cite{mikhailov} disorder-dependent length. 

\section{Summary}

We have studied double-layer quantum dot systems focusing 
on the MDD droplet states corresponding to the bulk incompressible
quantum Hall states at $\nu=1$ and $\nu=2$.  Ground states 
and collective excitations have been approximated using 
Hartree-Fock and time-dependent-Hartree-Fock approximations.
The Hartree-Fock $\nu=1$ MDD ground states have only symmetric
double-layer orbitals occupied and have spontaneous 
interlayer phase coherence.  The regime of stability of this
state has been estimated as a function of layer separation and confinement 
potential strength.  The broken symmetry in this state gives rise 
to low-energy excitations geometrically confined `Goldstone' 
collective excitations of the quantum dot.  The Hartree-Fock
$\nu=2$ MDD ground state has orbitals in both layers occupied.
The only low-energy collective excitations of this system
are localized near the edge of the system and correspond to
coupled magnetoplasmon excitations of bulk double-layer systems. 
We have used our microscopic calculations for quantum dots 
to derive expressions for the magnetoplasmon energies of 
bulk systems. 

\section{Acknowledgments}

JH thanks Nick Bonesteel and
Eduardo Miranda,  and AHM thanks Ulrich Z\" ulicke and 
Rudolf Haussmann for helpful discussions.
This work was supported by NSF grant DMR-9416906, Florida State E\&G
5024-010-02 and ONR N00014-93-0495.


\begin{references} 

\bibitem{doublelayer} See for example
J. Smoliner, E. Gornik and G. Weimann, Appl. Phys. Lett.
{\bf 52}, 2136 (1988); T.J. Gramila, J.P. Eisenstein, A.H. MacDonald,
L.N. Pfeiffer, and K.W. West, Phys. Rev. Lett. {\bf 66}, 1216 (1991);
Phys. Rev. B {\bf 47}, 12957 (1993); U. Sivan, P.M. Solomon,
and H. Shtrikman, Phys. Rev. Lett. {\bf 68}, 1196 (1992);
P.J. Price, Physica B {\bf 117}, 750 (1983);
H.C. Tso, P. Vasilopoulos, and F.M. Peeters, Phys. Rev. Lett.
{\bf 68}, 2516 (1992); A.-P. Jauho, and H. Smith, Phys. Rev.
B {\bf 47}, 4420 (1993); L. Zheng, and A.H. MacDonald,
Phys. Rev. B {\bf 48}, 8203 (1993); K. Flensberg and Ben
Yu-Kuang Hu, Phys. Rev. Lett. {\bf 73}, 3572 (1994);
J.P. Eisenstein, L.N. Pfeiffer, and K.W. West,
Phys. Rev. B {\bf 50}, 1760 (1994); J.P. Eisenstein, L.N. Pfeiffer, and
K.W. West, Phys. Rev. Lett. {\bf 68}, 674 (1992).

\bibitem{qdot}  For reviews see
U. Merkt, Advances in Solid State Physics, {\bf
30}, 77 (1990); Tapash Chakraborty,
Comments on Condensed Matter Physics {\bf 16}, 35 (1992);
M.A. Kastner, Rev. Mod. Phys. {\bf 64}, 849 (1992).

\bibitem{girvin1} S. M. Girvin and A. H. MacDonald, in {\it Novel
Quantum Liquids in Low-Dimensional Semiconductor Structures}, edited
by Sankar Das Sarma and Aron Pinxzuk (Wiley, New York, 1995); and
references therein.


\bibitem{murphy} S.Q. Murphy, J.P. Eisenstein, G.S. Boebinger,
L.N. Pfeiffer, and K.W. West, Phys. Rev. Lett. {\bf 72}, 728 (1994).

\bibitem{fertig1} H. A. Fertig, {\it Phys. Rev. B} {\bf 40}, 1989 (1087).

\bibitem{wen1} X. G. Wen and A. Zee, {\it Phys. Rev. Lett.} {\bf 69} 1992
(1811); X. G. Wen and A. Zee, {\it Phys. Rev. B} {\bf 47}, 1993 (2265).

\bibitem{macdonald3}A. H. MacDonald, P. M. Platzman and G. S. Boebinger, 
{\it Phys.  Rev. Lett.} {\bf 65}, (1990) 775.

\bibitem{yang1} Kun Yang, K. Moon, L. Zheng, A.H. MacDonald,
S.M. Girvin, D. Yoshioka, and Shou-Cheng Zhang,
Phys. Rev. Lett. {\bf 72}, 732 (1994).

\bibitem{moon1} K. Moon {\rm et al} {\it Phys. Rev. B} {\bf 51}, 1995 (5138).

\bibitem{aspec} P.L. McEuen {\it et al.}, Phys. Rev. Lett. {\bf 66}, 1926
(1991); R.C. Ashoori {\it et al.}, Phys. Rev. Lett. {\bf 71}, 613 (1993).

\bibitem{ajp} A. H. MacDonald, S. R. Eric Yang and M. D. Johnson,
{\it Aust. J. Phys.} {\bf 46}, (1993) 345. 

\bibitem{bsapology} Strictly speaking when we go beyond the 
Hartree-Fock approximation the broken symmetry occurs only in the 
$N \to \infty$ limit.  

\bibitem{pd} P. Hawvrylak, Phys. Rev. Lett. {\bf 71}, 3347 (1993);
J.J. Palacios {\it et al.}, Phys. Rev. B {\bf 50}, 5760 (1994);
S.R. Eric Yang, and M.D. Johnson, Phys. Rev. Lett. {\bf 71}, 3194 (1993).

\bibitem{erecons} C. de Chamon and X.-G. Wen, Phys. Rev. B {\bf 49}, 8227
(1994); O. Klein, C. de Chamon, D. Tang, D.M. Abusch-Magder, X.-G. Wen,
and M.A. Kastner, Phys. Rev. Lett. {\bf 74}, 785 (1995); 
J. Dempsey, B.Y. Gelfand, and B.I. Halperin, Phys. Rev. Lett. {\bf 70},
3639 (1993).

\bibitem{lqdots} G.S. Solomon, J.A. Trezza, A.F. Marshall, and J.S.
Harris Jr., Phys. Rev. Lett. {\bf 76}, 952 (1996) and work cited 
therein.

\bibitem{palacios} J. J. Palacios and P. Hawrylak, {\it Phys. Rev. B} {\bf 51}, 1995 (1769); Hiroshi Imamura, Peter A. Maksym, and Hideo Aoki,
preprint [cond-mat/9602120] (1996).

\bibitem{confpaper} J. Hu, E. Dagotto, and A.H. MacDonald,
{\it Physical Phenomena At High Magnetic Fields II}, edited by Z.~Fisk,
L. P. ~ Gor'kov, D. Meltzer and J. R. Schrieffer. (World Scientific, 1996) 

\bibitem{glazmanes} S.S. Nazin and V.B. Shikin, Zh. Eksp. Fiz. {\bf 85},
530 (1983) [Sov. Phys. JETP {\bf 58}, 210 (1983)]; V.B. Shikin, T.
Demel' and D. Heitmann, Zh. Eksp. Teor. Fiz. {\bf 96}, 1406 (1989) [Sov. Phys.
JETP {\bf 69}, 797 (1989)];
D.B. Chklovskii, B.I. Shklovskii, and 
L.I. Glazman, Phys. Rev. B {\bf 46}, 4026 (1992); 
M.M. Fogler, E.I. Levin, and B.I. Shklovskii, Phys. Rev. B
{\bf 49}, 13767 (1994)

\bibitem{edge} R.B. Laughlin, Phys. Rev. B {\bf 23},
5632 (1981); B.I. Halperin, Phys. Rev. B {\bf 25}, 2185 (1982);
A.H. MacDonald and P. Streda, Phys. Rev. B {\bf 29}, 1616 (1984);
M. Buttiker, Phys. Rev. B {\bf 38}, 9375 (1988);
X.G. Wen, Phys. Rev. B {\bf 41}, 12838 (1990);
D.H. Lee and X.G. Wen, Phys. Rev. Lett. {\bf 66}, 1765 (1991); X.G. Wen,
Phys. Rev. B {\bf 44}, 5708 (1991).
X.G. Wen, Int. J. Mod. Phys. {\bf B6}, 1711 (1992); A.H.  MacDonald,
Phys. Rev. Lett. {\bf 64}, 222 (1990).

\bibitem{macdonald2} A. H. MacDonald, {\it Surface Science} {\bf 229}, 
(1990) 1.

\bibitem{sondhikarlhede} S.L. Sondhi, A. Karlhede, and 
S.A. Kivelson, Bul. Am. Phys. Soc. {\bf 41}, 482 (1996).

\bibitem{empl} S.J. Allen, H.L. St\" ormer, and J.C.M. Hwang,
Phys. Rev. B {\bf 28}, 4875 (1983); D.B. Mast, A.J. Dahm,
and A.L. Fetter, Phys. Rev. Lett. {\bf 54}, 1706 (1985);
S.A. Govorkow {\it et al.}, Pis'ma Zh. Eksp. Teor. Fiz.
{\bf 44}, 380 (1986) [Sov. Phys. JETP Lett. {\bf 44}, 487 (1986)];
M. Wassermeier {\it et al.}, Phys. Rev. B {\bf 41}, 10287 (1990)
and work cited therein.

\bibitem{mikhailov} See V.A. Volkov and S.A. Mikhailov, Zh.
Eksp. Teor. Fiz {\bf 94}, 217 (1988) 
[Sov. Phys. JETP {\bf 67}, 1639 (1988)] and work cited therein.

\end{references}
\end{document}